# Realistic interpretation of a superposition state does not imply a mixture

Boon Leong Lan

School of Eng. & Sci., Monash University, 2 Jalan Kolej, 46150 Petaling Jaya, Selangor, Malaysia

**Abstract**

Contrary to previous claims, it is shown that, for an ensemble of either single-particle systems or multi-particle systems, the realistic interpretation of a superposition state that mathematically describes the ensemble does not imply that the ensemble is a mixture. Therefore it cannot be argued that the realistic interpretation is wrong on the basis that some predictions derived from the mixture are different from the corresponding predictions derived from the superposition state.

It has been claimed, in [1-3] for example, that the realistic interpretation of a superposition state, $\sum_k a_k |\varphi_k\rangle$, which mathematically describes an ensemble of either single-particle systems [1] or multi-particle systems [2,3] implies that the ensemble is a mixture. In a mixture, $|a_k|^2 \times 100$ percent of the systems in the ensemble are each mathematically described by the state $|\varphi_k\rangle$. There are instances [1-5] where the prediction that is derived from a mixture is different from the prediction derived from a superposition state. This has led to the conclusion [3] that the realistic interpretation of a superposition state is incorrect. In this paper, I will show however that, in both the single-particle and multi-particle case, the realistic interpretation of a superposition state does not imply a mixture, contrary to previous claims.

We begin with the single-particle case. Consider, for simplicity, an ensemble of single-particle systems that is mathematically described by this superposition state:

$$\psi = \sqrt{\frac{2}{3}}|0\rangle + \frac{1}{\sqrt{3}}|1\rangle. \tag{1}$$

According to the realistic interpretation of $\psi$, prior to measurement, two-thirds of the ensemble is actually in state $|0\rangle$, and the remainder one-third of the ensemble is actually in state $|1\rangle$. First, for simplicity, lets consider the case where there are only three single-particle systems, S1, S2 and S3, in the ensemble. The first possibility is that, prior to measurement, S1 is actually in state $|0\rangle$, S2 is actually in state $|0\rangle$, and S3 is actually in state $|1\rangle$. The second possibility is that, prior to measurement, S1 is actually in state $|0\rangle$, S2 is actually in state $|1\rangle$, and S3 is actually in state $|0\rangle$. The third and final possibility is that, prior to measurement, S1 is actually in state $|1\rangle$, S2 is actually in state $|0\rangle$, and S3 is actually in state $|0\rangle$. (In other words, although two of the three systems in the ensemble are actually in state $|0\rangle$ and one of the three systems in the ensemble is actually in state $|1\rangle$ prior to measurement according to a realist, it is not known exactly which two systems in the ensemble are actually in state $|0\rangle$ and exactly which system in the ensemble is actually in state $|1\rangle$ prior to measurement.) From these three possibilities, it follows that, prior to measurement, S1 is actually either in state $|0\rangle$ with probability 2/3 or in state $|1\rangle$ with probability 1/3, S2 is also actually either in state $|0\rangle$ with probability 2/3 or in state $|1\rangle$ with probability 1/3, and S3 is also actually either in state $|0\rangle$ with probability 2/3 or in state $|1\rangle$ with probability 1/3. Thus, from a realistic viewpoint, S1, S2 and S3 are *each* mathematically described also by the superposition state $\psi$ in Eq. (1), consistent with the fact that we have an ensemble of systems where each member of the ensemble is mathematically described by $\psi$. The same conclusion holds for each system for an arbitrary number of systems in the ensemble. Hence, the ensemble is not a mixture where two-thirds of the systems in the ensemble are *each* mathematically described by the state $|0\rangle$ and the other one-third of the systems in the ensemble are *each* mathematically described by the state $|1\rangle$. A similar argument for any other superposition state that mathematically describes an ensemble of single-particle systems will lead to the same conclusion that the realistic interpretation of the state does not imply that the ensemble is a mixture.

The argument for the multi-particle case is similar to the one for the single-particle case and I will illustrate it for an ensemble of two-particle systems that is mathematically described by this superposition state:

$$\phi = \frac{1}{\sqrt{2}}|0\rangle|0\rangle + \frac{1}{\sqrt{2}}|1\rangle|1\rangle. \qquad (2)$$

The realistic interpretation of $\phi$ implies that, prior to measurement, half of the ensemble is actually in state $|0\rangle|0\rangle$ and the other half of the ensemble is actually in state $|1\rangle|1\rangle$. For simplicity, lets first consider an ensemble of only two two-particle systems, T1 and T2. In this case, there are only two possibilities: prior to measurement, either T1 is actually in state $|0\rangle|0\rangle$ (i.e., each of the two particles of system T1 is actually in state $|0\rangle$) and T2 is actually in state $|1\rangle|1\rangle$ (i.e., each of the two particles of system T2 is actually in state $|1\rangle$), or, T1 is actually in state $|1\rangle|1\rangle$ and T2 is actually in state $|0\rangle|0\rangle$. (In other words, although one of the two systems in the ensemble is actually in state $|0\rangle|0\rangle$ and the other system is actually in state $|1\rangle|1\rangle$ prior to measurement from a realistic viewpoint, it is not known exactly which system is actually in state $|0\rangle|0\rangle$ and exactly which system is actually in state $|1\rangle|1\rangle$ prior to measurement.) These two possibilities imply that, prior to measurement, T1 is actually either in state $|0\rangle|0\rangle$ or in state $|1\rangle|1\rangle$ with equal probability of $1/2$, and T2 is also actually either in state $|0\rangle|0\rangle$ or in state $|1\rangle|1\rangle$ with equal probability of $1/2$. So, from a realistic viewpoint, T1 and T2 are *each* mathematically described also by the superposition state $\phi$ in Eq. (2). Likewise, for an ensemble of an arbitrary number of systems, each member of the ensemble is, from a

realistic viewpoint, also mathematically described by $\phi$. Therefore, the realistic interpretation of the superposition state $\phi$ does not imply that the ensemble is a mixture where half of the systems in the ensemble are *each* mathematically described by the state $|0\rangle|0\rangle$ and the other half of the systems in ensemble are *each* mathematically described by the state $|1\rangle|1\rangle$.

Because the realistic interpretation of a superposition state that mathematically describes an ensemble does not imply that the ensemble is a mixture, it cannot be argued, as Furry [3] did for example, that the realistic interpretation is wrong on the basis that some predictions derived from the mixture are different from the corresponding predictions derived from the superposition state.